\documentclass[letterpaper]{article}
\pdfoutput=1

\usepackage{graphicx}
\usepackage{hyperref}

\usepackage[margin=2.5cm]{geometry}
\usepackage[font=sf]{caption}

\long\def\symbolfootnote[#1]#2{\begingroup%
\def\thefootnote{\fnsymbol{footnote}}\footnote[#1]{#2}\endgroup} 

\begin{document}

\begin{center}

{\LARGE Strong coupling between single-electron
 tunneling and  nano-mechanical motion}

\vspace{2em}

G. A. Steele$^1$, 
A. K. H\"{u}ttel$^{1,*}$\symbolfootnote[0]{
$^*$Present address: Institute for Experimental
  and Applied Physics, University of Regensburg, 93040 Regensburg, Germany},
B. Witkamp$^1$, M. Poot$^1$, H. B. Meerwaldt$^1$,
L. P. Kouwenhoven$^1$ \\ and H. S. J. van der Zant$^1$ 

{\em $^1$Kavli Institute of NanoScience, Delft University of Technology, PO Box 
5046, 2600 GA, Delft, The Netherlands.}

\end{center}

\vspace{2em}

\begin{abstract}
Nanoscale resonators that oscillate at high frequencies are useful in
many measurement applications. We studied a high-quality mechanical
resonator made from a suspended carbon nanotube driven into motion by
applying a periodic radio frequency potential using a nearby
antenna. Single-electron charge fluctuations created periodic
modulations of the mechanical resonance frequency. A quality factor
exceeding $10^5$ allows the detection of a shift in resonance frequency
caused by the addition of a single-electron charge on the
nanotube. Additional evidence for the strong coupling of mechanical
motion and electron tunneling is provided by an energy transfer to the
electrons causing mechanical damping and unusual nonlinear
behavior. We also discovered that a direct current through the
nanotube spontaneously drives the mechanical resonator, exerting a
force that is coherent with the high-frequency resonant mechanical
motion.
\end{abstract}
\vfill
\pagebreak

Nanomechanical systems \cite{Craighead2000Nanoelectromechanical,
Ekinci2005Nanoelectromechanical} have promising applications, such as
ultra-sensitive mass detection
\cite{Ekinci2004Ultrasensitive,Lassagne2008Ultrasensitive,Chiu2008AtomicScale}.
The combination of a high resonance frequency and a small mass also
makes nanomechanical resonators attractive for a fundamental study of
mechanical motion in the quantum limit
\cite{Knobel2003Nanometrescale,Lahaye2004Approaching,
Naik2006Cooling,Schwab2005Putting}.  For a successful observation of
quantum motion of a macroscopic object, a high-frequency nanoscale
resonator must have low dissipation (which implies a high
quality-factor Q), and a sensitive detector with minimum back-action
(i.e. quantum limited)
\cite{Caves1980Measurement,Regal2008Measuring}. Here, we demonstrate a
dramatic backaction that strongly couples a quantum dot detector to
the resonator dynamics of a carbon nanotube, and which, in the limit
of strong feedback, spontaneously excites large amplitude resonant
mechanical motion.

Nanomechanical resonators have been realized by etching down larger
structures. In small devices, however, surfaces effects impose a limit
on the quality-factor
\cite{Ekinci2005Nanoelectromechanical}. Alternatively, suspended
carbon nanotubes can be used to avoid surface damage from the
(etching) fabrication process. We recently developed a mechanical
resonator based on an ultra-clean carbon nanotube with high resonance
frequencies of several 100 MHz and a Q exceeding $10^5$
\cite{Huettel2009Carbon}. Here, we exploit this resonator to explore a
strong coupling regime between single electron tunneling and
nanomechanical motion. We followed the pioneering approaches in which
aluminium single electron transistors were used as position detectors
\cite{Knobel2003Nanometrescale,Lahaye2004Approaching,Naik2006Cooling}
and AFM cantilevers as resonators
\cite{Woodside2002Scanned,Zhu2005Frequency, Stomp2005Detection};
however, our experiment is in the limit of much stronger
electro-mechanical coupling, achieved by embedding a quantum dot
detector in the nanomechanical resonator itself.

Our device consists of a nanotube suspended across a trench that makes
electrical contact to two metal electrodes (Fig.\ 1). Electrons are
confined in the nanotube by Schottky barriers at the Pt metal
contacts, forming a quantum dot in the suspended segment. The nanotube
growth is the last step in the fabrication process, yielding
ultra-clean devices \cite{Steele2009Tunable}, as demonstrated by the
four-fold shell-filling of the Coulomb peaks (Fig.\ 1C). All
measurements were performed at a temperature of 20 mK with an electron
temperature of $\sim$80 mK.

We actuate the resonator with a nearby antenna, and detect the
resonator motion by its influence on the d.c.\ current through the
nanotube. The inset to Fig.\ 1D shows a peak in the current at the
resonance frequency, which we have identified as a bending-mode
mechanical resonance of the nanotube \cite{Huettel2009Carbon}. The
Q-factor typically exceeds $10^5$, which is an increase of more than
two orders of magnitude compared to previous nanotube studies
\cite{Sazonova2004Tunable,
Witkamp2006BendingMode,Lassagne2008Ultrasensitive}. The resonance
frequency is tuned by more than a factor of 2 with the gate voltage
(Fig.\ 1D). Here, the electric field from the gate voltage pulls the
nanotube toward it, and the subsequent lengthening of the nanotube
induces more tension, similar to the tuning of a guitar string
\cite{Sazonova2004Tunable}.

Our detection signal results from a change in gate capacitance,
$\Delta C_g$, during a displacement of the nanotube. This changes the
effective quantum dot potential and, if positioned initially beside a
Coulomb peak (Fig.\ 1C), can move it onto the peak, thereby increasing
the current. For a nanotube oscillating on resonance, the effective
potential oscillates, and the non-linearity of Coulomb blockade allows
it to be rectified to a detectable d.c.\ current.

The narrow linewidth of the resonance peak due to the high Q-factor
provides an unprecedented sensitive probe for studying nanomechanical
motion. We first show the influence of a single electron on the
resonance frequency, $f_0$. The Coulomb oscillations in Fig.\ 2A are
caused by single electron tunneling giving rise to current peaks, and
Coulomb blockade fixes the electron number in the valleys.  From
valley to valley, the electron number changes by one. Fig.\ 2B shows
the mechanical resonance signal recorded at the same time. Overall, a
more negative gate voltage (right to left) increases the total charge
on the nanotube, increasing the tension. This process stiffens the mechanical
spring constant and increases the resonance frequency. Linear
stiffening occurs in the Coulomb valleys (indicated with dashed
lines), whereas at Coulomb peaks, a peculiar softening occurs, visible
as dips in $f_0$.

We first focus on the change in resonance frequency caused by the
addition of one electron, which is measured as offsets of about 0.1
MHz between the dashed lines. This shift from single electron tuning,
predicted in \cite{Sapmaz2003Carbon}, is about 20 times our linewidth
and thus clearly resolvable.  Because we compare valleys with a fixed
electron number, this single electron tuning comes from a change in a
static force on the nanotube. The (electro-) static force is
proportional to the square of the charge on the nanotube and thus
adding one electron charge results, here, in a detectable shift in the
mechanical resonance \cite{Sapmaz2003Carbon}. The shifts from single
electron tuning can be as large as 0.5 MHz, more than 100 times the
line width \cite{supp}.

Next we focus on the dips in resonance frequency that occur at the
Coulomb peaks. The current at the Coulomb peaks is carried by single
electron tunneling, meaning that one electron tunnels off the nanotube
before the next electron can enter the tube. The charge on the
nanotube thus fluctuates by exactly one electron charge, $e$, with a
time dynamics than can be accounted for in detail by the theory of
Coulomb blockade \cite{Beenakker1991Theory}. The average rate,
$\Gamma$, at which an electron moves across the tube can be read off
from the current $I = e\Gamma$ (1.6 pA corresponds to a 10 MHz rate).
Moving the gate voltage off or on a Coulomb peak, we can tune the rate
from the regime $\Gamma \sim f_0$ to $\Gamma \gg f_0$ and explore the
different effects on the mechanical resonance.

In Figs.\ 2A,B the Coulomb peak values of $\sim$ 8 nA yield $\Gamma
\sim 300 f_0$, the regime of many single electron tunneling events per
mechanical oscillation. In addition to the static force and the radio
frequency (RF) oscillating driving force, single electron tunneling
now exerts a time-fluctuating, dynamic force on the mechanical
resonator. We observe that this dynamic force causes softening, giving
dips in the resonance frequency. The single electron charge
fluctuations do not simply smooth the stepwise transition from the
static single electron tuning shifts. Strikingly, fluctuations instead
caused dips in the resonant frequency up to an order of magnitude
greater than the single electron tuning shifts. As shown in
\cite{Woodside2002Scanned,brinkthesis} and discussed in detail in the
supporting online material \cite{supp}, the dynamic force modifies the
nanotube†¢s spring constant, $k$, resulting in a softening of the
mechanical resonance. The shape of the frequency dip can be altered by
applying a finite bias, $V_{sd}$, across the nanotube. Starting from
deep and narrow at small $V_{sd}$ = 0.5 mV, the dip becomes shallower
and broader with increasing $V_{sd}$. This dip-shape largely resembles
the broadening of Coulomb blockade peaks that occurs with increasing
$V_{sd}$. We thus conclude from Fig.\ 2 that the single electron
tuning oscillations are a mechanical effect that is a direct
consequence of single electron tunneling oscillations.

Besides softening, the charge fluctuations also provide a channel for
dissipation of mechanical energy.  Fig.\ 3A shows the resonance dip
for small RF power with frequency traces in Fig.\ 3B. In the Coulomb
valleys, tunneling is suppressed ($\Gamma \sim f_0$), damping of the
mechanical motion is minimized, and we observe the highest
Q-factors. On a Coulomb peak, charge fluctuations are maximal
($\Gamma \gg f_0$), and the Q-factor decreases to a few
thousand. These results explicitly show that detector backaction can
cause significant mechanical damping. The underlying mechanism for the
damping is an energy transfer occasionally occurring when a
current-carrying electron is pushed up to a higher (electrostatic)
energy by the nanotube motion before tunneling out of the dot. This
gain in potential energy is later dissipated in the drain contact.

If we drive the system at higher RF powers (Fig.\ 3C,D) we observe an
asymmetric resonance peak, along with distinct hysteresis between
upward and downward frequency sweeps. Theoretically this marks the
onset of non-linear terms in the equation of motion, such as in the
well-studied Duffing oscillator \cite{cleland_book,Nayfeh}. The spring
constant, $k$, is modified by a large oscillation amplitude, $x$,
which is accounted for by replacing $k$ with $(k+\alpha x^2)$. The
time-averaged spring constant increases if $\alpha > 0$, which is
accompanied by a sharp edge at the high frequency side of the peak;
vice versa for $\alpha < 0$.  In addition to the overall softening of
$k$ yielding the frequency dips of Fig.\ 2, the fluctuating charge on
the dot also changes $\alpha$, giving a softening spring ($\alpha <
0$) outside of the frequency dip (Coulomb valleys), and a hardening
spring ($\alpha > 0$) inside the frequency dip (Coulomb peaks), shown
in Fig.\ 3. The sign of $\alpha$ follows the curvature of $f_0(V_g)$
induced by the fluctuating electron force, giving a change in sign at
the inflection point of the frequency dip. Interestingly,
non-linearity from the single electron force in our device dominates,
and is much stronger than that from the mechanical deformation
\cite{supp}.

Figs. 3E,F show the regime of further enhanced RF driving. The
non-linearity is now no longer a perturbation of the spring constant,
but instead gives sharp peaks in the lineshape and switching between
several different metastable modes (see further data in supplementary
material \cite{supp}). At this strong driving, we observe highly
structured nonlinear mechanical behavior that arises from the coupling
of the resonator motion to the quantum dot.

In figure 3, we studied non-linear coupling between the quantum dot
and the mechanical resonator by applying a large RF driving force at a
small Vsd. In figure 4, we consider a small or absent RF driving force
and now apply a large Vsd across the quantum dot. Fig. 4A shows a
standard Coulomb blockade measurement of the quantum dot. Mechanical
effects in Coulomb diamonds have been studied before in the form of
phonon sidebands of electronic transitions
\cite{Park2000Nanomechanical,
Sapmaz2006Tunneling,Zwanenburg2009Spin,Leturcq2009FranckCondon}. Shown
in the data of figure 4 are reproducible ridges of positive and
negative spikes in the differential conductance as indicated by
arrows. This instability has been seen in all 12 measured devices with
clean suspended nanotubes and never in non-suspended devices. Fig.\ 4B
and C shows such ridges in a second device, visible both as spikes in
the differential conductance (Fig.\ 4B), and as discrete jumps in the
current (Fig.\ 4C).  The barriers in device 2 were highly tunable: we
found that the switch-ridge could be suppressed by reducing the tunnel
coupling to the source-drain leads, thereby decreasing the
current. The instability disappears roughly when the tunnel rate is
decreased below the mechanical resonance frequency (see supporting
online material)\cite{supp}.

In a model predicting such instabilities \cite{Usmani2007Strong},
positive feedback from single electron tunneling excites the
mechanical resonator into a large amplitude oscillation.  The theory
predicts a characteristic shape of the switch-ridges and the
suppression of the ridges for $\Gamma \sim$ $f_0$, in striking
agreement with our observations.  Such feedback also requires a very
high Q, which may explain why it has not been observed in previous
suspended quantum dot devices
\cite{Sapmaz2006Tunneling,Leturcq2009FranckCondon}.  If the required
positive feedback is present, however, it should also have a
mechanical signature: such a signature is demonstrated in Fig.\
4E. The RF-driven mechanical resonance experiences a dramatic
perturbation triggered by the switch-ridge discontinuities in the
Coulomb peak current shown in Fig.\ 4D. At the position of the switch,
the resonance peak shows a sudden departure from the expected
frequency dip (dashed line), and becomes strongly asymmetric and
broad, as if driven by a much higher RF power. This is indeed the
case, but the driving power is now provided by an internal source:
because of the strong feedback, the random fluctuating force from
single electron tunneling becomes a driving force coherent with the
mechanical oscillation. Remarkably, the d.c.\ current through the
quantum dot can be used both to detect the high-frequency resonance
and, in the case of strong feedback, directly excite resonant
mechanical motion.

%\bibliography{paper}{}

\begin{thebibliography}{10}

\bibitem{Craighead2000Nanoelectromechanical}
H.~G. Craighead, {\it Science\/}
  \href{http://dx.doi.org/10.1126/science.290.5496.1532} {{\bf 290}, 1532}
  (2000).

\bibitem{Ekinci2005Nanoelectromechanical}
K.~L. Ekinci, M.~L. Roukes, {\it Rev. Sci. Inst.\/}
  \href{http://scitation.aip.org/getabs/servlet/GetabsServlet?prog=normal\&id=%
RSINAK000076000006061101000001\&idtype=cvips\&gifs=yes} {{\bf 76}, 061101}
  (2005).

\bibitem{Ekinci2004Ultrasensitive}
K.~L. Ekinci, X.~M.~H. Huang, M.~L. Roukes, {\it Appl. Phys. Lett.\/}
  \href{http://scitation.aip.org/getabs/servlet/GetabsServlet?prog=normal\&id=%
APPLAB000084000022004469000001\&idtype=cvips\&gifs=yes} {{\bf 84}, 4469}
  (2004).

\bibitem{Lassagne2008Ultrasensitive}
B.~Lassagne, D.~Garcia-Sanchez, A.~Aguasca, A.~Bachtold, {\it Nano Lett.\/}
  \href{http://dx.doi.org/10.1021/nl801982v} {{\bf 8}, 3735} (2008).

\bibitem{Chiu2008AtomicScale}
H.-Y. Chiu, P.~Hung, H.~W. Postma, M.~Bockrath, {\it Nano Lett.\/}
  \href{http://dx.doi.org/10.1021/nl802181c} {{\bf 8}, 4342} (2008).

\bibitem{Knobel2003Nanometrescale}
R.~G. Knobel, A.~N. Cleland, {\it Nature\/}
  \href{http://dx.doi.org/10.1038/nature01773} {{\bf 424}, 291} (2003).

\bibitem{Lahaye2004Approaching}
M.~D. Lahaye, O.~Buu, B.~Camarota, K.~C. Schwab, {\it Science\/}
  \href{http://dx.doi.org/10.1126/science.1094419} {{\bf 304}, 74} (2004).

\bibitem{Naik2006Cooling}
A.~Naik, {\it et~al.\/}, {\it Nature\/}
  \href{http://dx.doi.org/10.1038/nature05027} {{\bf 443}, 193} (2006).

\bibitem{Schwab2005Putting}
K.~C. Schwab, M.~L. Roukes, {\it Phys. Today\/}
  \href{http://dx.doi.org/10.1063/1.2012461} {{\bf 58}, 36} (2005).

\bibitem{Caves1980Measurement}
C.~M. Caves, K.~S. Thorne, R.~W. Drever, V.~D. Sandberg, M.~Zimmermann, {\it
  Rev. Mod. Phys.\/} \href{http://dx.doi.org/10.1103/RevModPhys.52.341} {{\bf
  52}, 341} (1980).

\bibitem{Regal2008Measuring}
C.~A. Regal, J.~D. Teufel, K.~W. Lehnert, {\it Nature Phys.\/}
  \href{http://dx.doi.org/10.1038/nphys974} {{\bf 4}, 555} (2008).

\bibitem{Huettel2009Carbon}
A.~K. Huettel, {\it et~al.\/}, {\it Nano Letters\/}
  \href{http://dx.doi.org/10.1021/nl900612h} {{\bf 9}, 2547} (2009).

\bibitem{Woodside2002Scanned}
M.~T. Woodside, P.~L. Mc{E}uen, {\it Science\/}
  \href{http://dx.doi.org/10.1126/science.1069923} {{\bf 296}, 1098} (2002).

\bibitem{Zhu2005Frequency}
J.~Zhu, M.~Brink, P.~L. Mc{E}uen, {\it Appl. Phys. Lett.\/}
  \href{http://scitation.aip.org/getabs/servlet/GetabsServlet?prog=normal\&id=%
APPLAB000087000024242102000001\&idtype=cvips\&gifs=yes} {{\bf 87}} (2005).

\bibitem{Stomp2005Detection}
R.~Stomp, {\it et~al.\/}, {\it Phys. Rev. Lett.\/}
  \href{http://dx.doi.org/10.1103/PhysRevLett.94.056802} {{\bf 94}, 056802}
  (2005).

\bibitem{Steele2009Tunable}
G.~A. Steele, G.~Gotz, L.~P. Kouwenhoven, {\it Nature Nano.\/}
  \href{http://dx.doi.org/10.1038/nnano.2009.71} {{\bf 4}, 363} (2009).

\bibitem{Sazonova2004Tunable}
V.~Sazonova, {\it et~al.\/}, {\it Nature\/}
  \href{http://dx.doi.org/10.1038/nature02905} {{\bf 431}, 284} (2004).

\bibitem{Witkamp2006BendingMode}
B.~Witkamp, M.~Poot, H.~S.~J. van~der Zant, {\it Nano Lett.\/}
  \href{http://dx.doi.org/10.1021/nl062206p} {{\bf 6}, 2904} (2006).

\bibitem{Sapmaz2003Carbon}
S.~Sapmaz, Y.~Blanter, L.~Gurevich, H.~S.~J. van~der Zant, {\it Phys. Rev. B\/}
  \href{http://dx.doi.org/10.1103/PhysRevB.67.235414} {{\bf 67}, 235414}
  (2003).

\bibitem{supp}
Supporting online material available on {\em Science} Online.

\bibitem{Beenakker1991Theory}
C.~W.~J. Beenakker, {\it Phys. Rev. B\/}
  \href{http://dx.doi.org/10.1103/PhysRevB.44.1646} {{\bf 44}, 1646} (1991).

\bibitem{brinkthesis}
M. Brink, thesis, Cornell Univerity (2007).

\bibitem{cleland_book}
A.~Cleland, {\it Foundations of Nanomechanics\/} (Springer-Verlag, 2002).

\bibitem{Nayfeh}
A.~H. Nayfeh, D.~T. Mook, {\it Nonlinear Oscillations\/} (Wiley, 1979).

\bibitem{Park2000Nanomechanical}
H.~Park, {\it et~al.\/}, {\it Nature\/}
  \href{http://dx.doi.org/10.1038/35024031} {{\bf 407}, 57} (2000).

\bibitem{Sapmaz2006Tunneling}
S.~Sapmaz, J.~P. Herrero, Y.~M. Blanter, C.~Dekker, H.~S.~J. van~der Zant, {\it
  Phys. Rev. Lett.\/} \href{http://dx.doi.org/10.1103/PhysRevLett.96.026801}
  {{\bf 96}} (2006).

\bibitem{Zwanenburg2009Spin}
F.~A. Zwanenburg, C.~E. van Rijmenam, Y.~Fang, C.~M. Lieber, L.~P. Kouwenhoven,
  {\it Nano Lett.\/} \href{http://dx.doi.org/10.1021/nl803440s} {{\bf 9}, 1071}
  (2009).

\bibitem{Leturcq2009FranckCondon}
R.~Leturcq, {\it et~al.\/}, {\it Nature Phys.\/}
  \href{http://dx.doi.org/10.1038/nphys1234} {{\bf 5}, 327} (2009).

\bibitem{Usmani2007Strong}
O.~Usmani, Y.~M. Blanter, Y.~V. Nazarov, {\it Phys. Rev. B\/}
  \href{http://scitation.aip.org/getabs/servlet/GetabsServlet?prog=normal\&id=%
PRBMDO000075000019195312000001\&idtype=cvips\&gifs=yes} {{\bf 75}, 195312}
  (2007).

\end{thebibliography}

\setlength{\parindent}{0pt}

\vspace{2em} 

{\bf \large Acknowledgments}

We thank Y. M. Blanter and Y. V. Nazarov for helpful discussions. This
work was supported by the Dutch Organization for Fundamental Research
on Matter (FOM), the Netherlands Organization for Scientific Research
(NWO), the Nanotechnology Network
Netherlands (NanoNed), and the Japan Science and Technology Agency
International Cooperative Research Project (JST-ICORP).

\vspace{2em}

\pagebreak

\begin{figure}
\begin{center}
\includegraphics[width=3.25in]{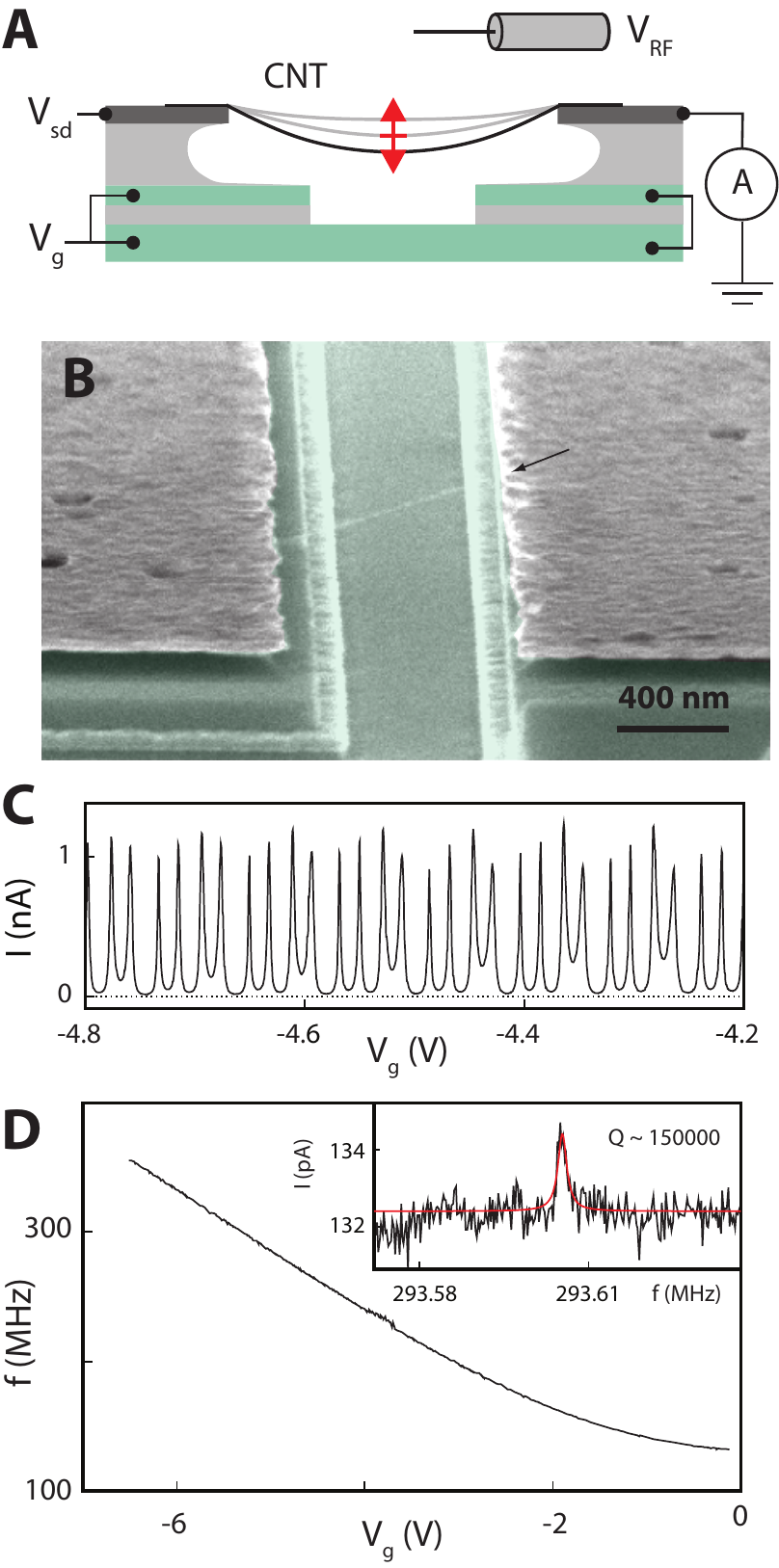}
\end{center}
\caption{A high quality-factor nanotube mechanical resonator with an
embedded quantum dot. (A) Device layout. A suspended carbon nanotube is
excited into mechanical motion by applying an a.c.\ voltage to a nearby
antenna. A d.c.\ current through the nanotube detects the
motion. $V_{RF}$, radio frequency voltage; CNT, carbon nanotube. (B)
SEM image of a typical device. (C) A quantum dot, formed between
Schottky barriers at the metal contacts, displays 4-fold shell filling
of holes. (D) (Inset) The mechanical resonance induces a corresponding
resonance in the d.c.\ current which can have a narrow linewidth with
quality-factors up to 150 000. (Main plot) The resonance frequency can
be tuned using a tensioning force from the d.c.\ voltage on the gate.}
\end{figure}

\begin{figure}
\begin{center}
\includegraphics{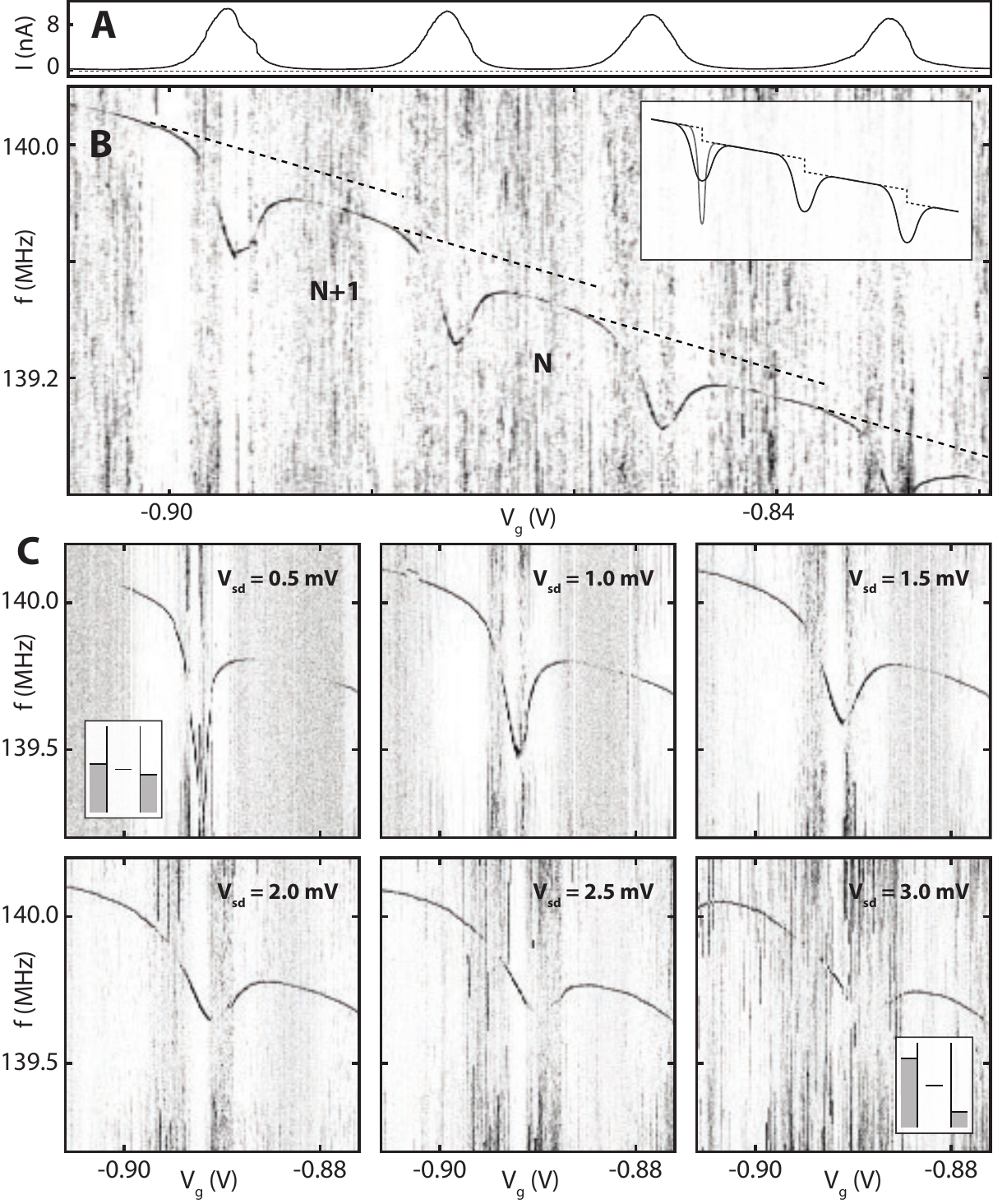}
\end{center}
\caption{Single electron tuning. (A) Nanotube current vs. gate voltage
showing single electron tunneling at the peaks and Coulomb blockade in
the valleys. This curve is taken from (B) at f = 138.8 MHz. (B)
Normalized resonance signal $\Delta I/\Delta I_{peak}$ (see supporting
online material) vs. RF frequency and gate voltage ($V_{sd}$ = 1.5
mV). The tuned mechanical resonance shows up as the darker curve with
dips at the Coulomb peaks. The offsets between dashed lines indicate
the frequency shift due to the addition of one electron to the
nanotube. The resonance frequency also shows dips caused by a
softening of the spring constant due to single electron charge
fluctuations. N, number of holes on the quantum dot. (Inset) The
expected resonance behavior (see text). (C) Zoom on one frequency dip
for various source-drain voltages ($V_{sd}$) showing dip broadening
for increasing $V_{sd}$. (Insets) Energy diagrams for small and large
$V_{sd}$.}
\end{figure}

\begin{figure}
\begin{center}
\includegraphics[width=6.5in]{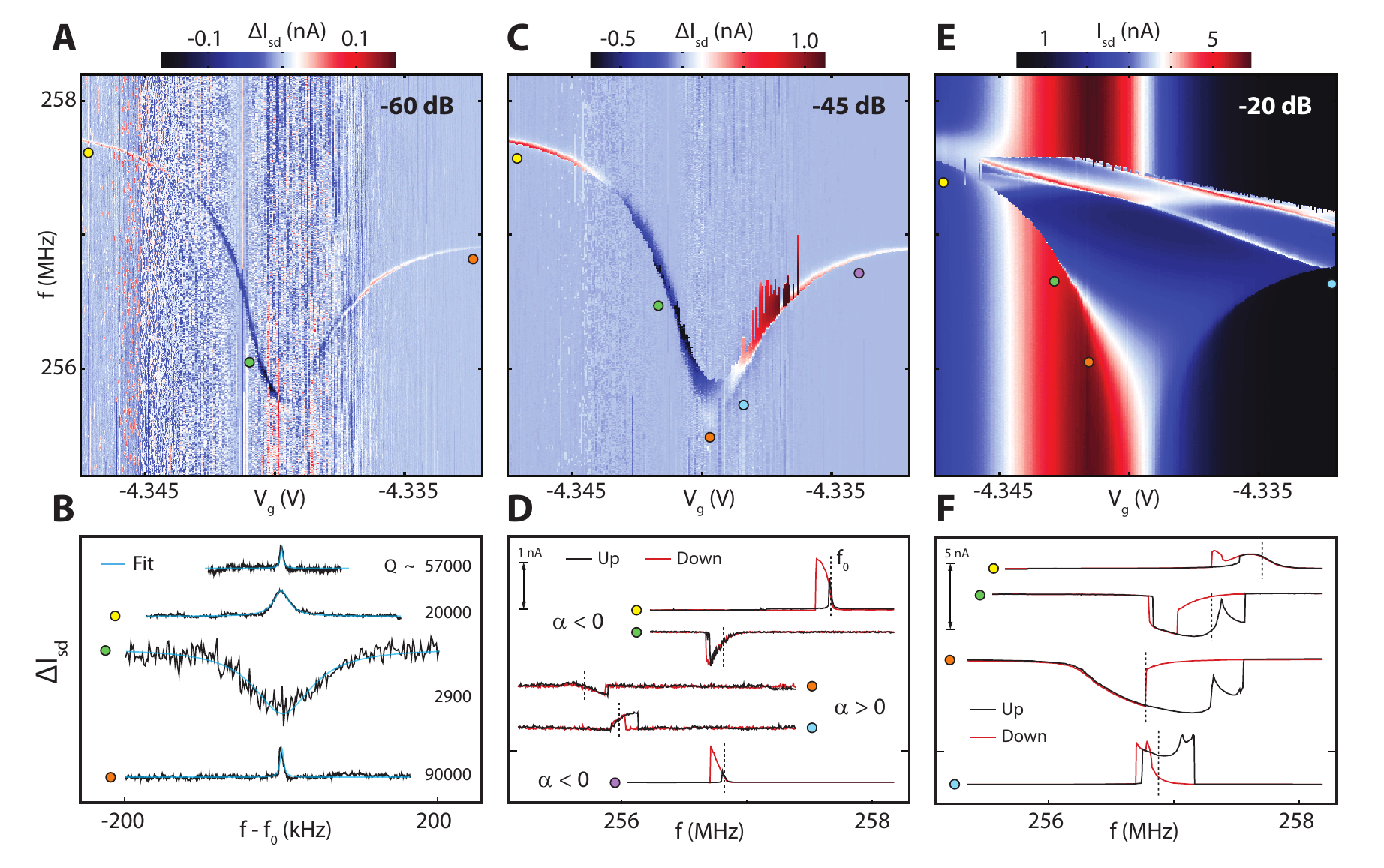}
\end{center}
\caption{Lineshapes of the mechanical resonance from linear to
non-linear driving regimes. (A) Detector current ($\Delta I$)
vs. frequency and gate-voltage at RF excitation power of -60 dB as the
gate voltage is swept through one Coulomb peak. (B) Fits of the
resonance to a squared Lorentzian lineshape at different gate voltages
\cite{Huettel2009Carbon}. The RF power for each trace is adjusted to
stay in the linear driving regime (-75,-64,-52, and -77 dB top to
bottom). Traces are taken at the positions indicated by colored
circles (aside from the top trace which is taken at $V_g$ = -4.35
V). (C) At -45 dB, the resonance has an asymmetric lineshape with one
sharp edge, see linecuts (D), typical for a non-linear oscillator
\cite{cleland_book,Nayfeh}. Dashed lines in (D) and (F) indicate
resonance frequency $f_0$ at low powers. (E) and (F) At even higher
driving powers (-20 dB), the mechanical resonator displays sharp
sub-peaks and several jumps in amplitude when switching between
different stable modes. (C) and (E) are taken in the upwards sweep
direction (downwards sweeps shown in supporting online material).}
\end{figure}

\begin{figure}
\begin{center}
\includegraphics[width=5.5in]{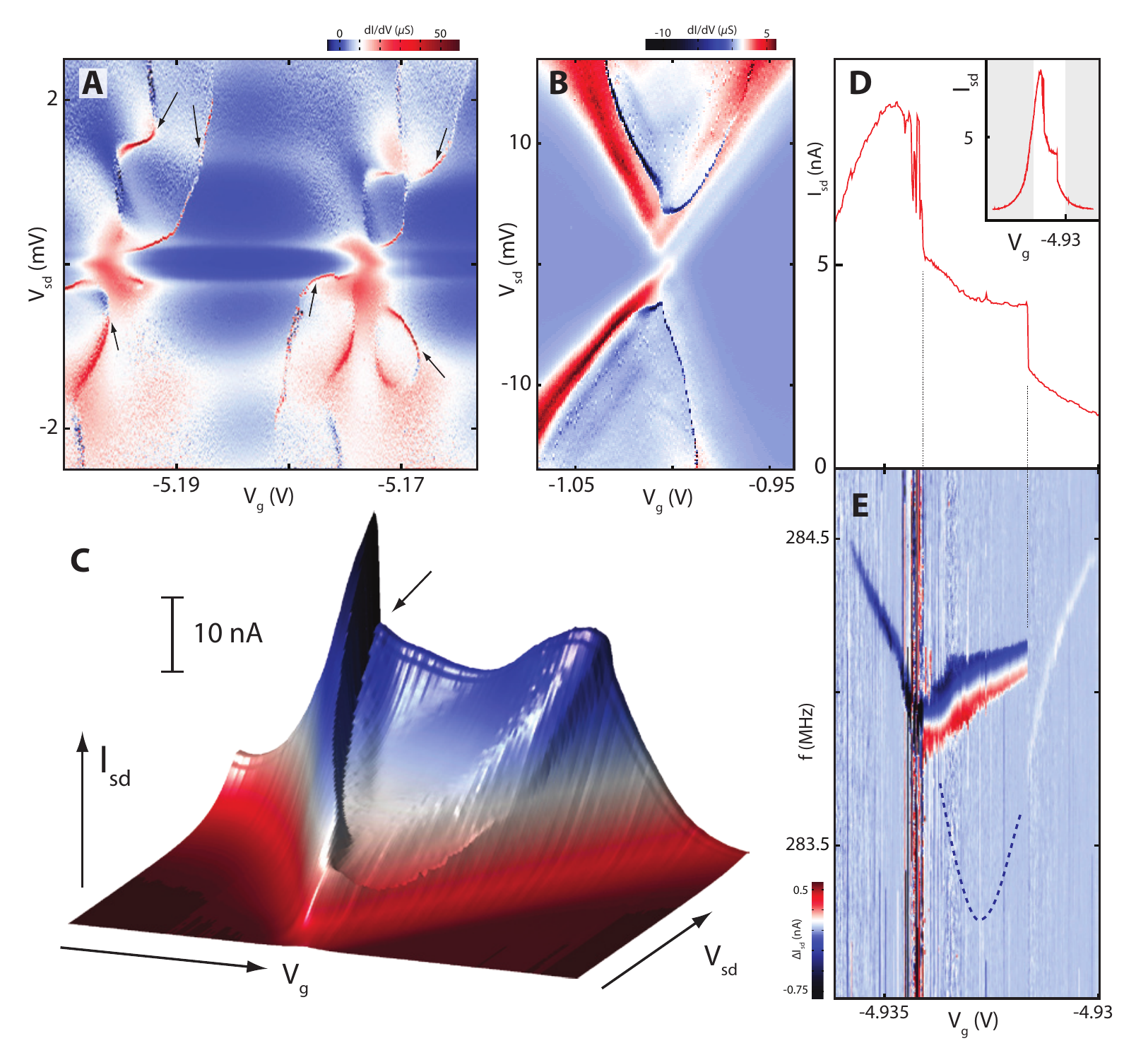}
\end{center}
\caption{Spontaneous driving of the mechanical resonance by
single-electron tunneling. (A) Differential conductance, $dI/dV_{sd}$,
showing ridges of sharp positive (deep red) and negative (deep blue)
spikes (arrows). (B) Similar ridges measured on device 2 (trench width
= 430 nm) in the few-hole regime (4hole to 3hole transition).  Spikes
in $dI/dV_{sd}$ appear as step-like ridges in current: (C) shows the
data from the upper half of (B), but now as a 3D current plot. Note
that the ridges are entirely reproducible. (D) Inset: Coulomb peak at
$V_{sd}$ = 0.5 mV showing large switching-steps. Main plot: zoom-in on
data from inset. (E) RF driven mechanical resonance measured for the
same Coulomb peak in (D) at a driving power of -50 dB. Outside the
``switch-region'', the resonance has a narrow lineshape and follows
the softening-dip from Figs. 2,3. At the first switch, the resonance
position departs from the expected position (indicated by dashed
line). The mechanical signal is strongly enhanced in amplitude and
displays a broad asymmetric lineshape. At the second switch, the
resonance returns to the frequency and narrow lineshape expected at
these powers.}
\end{figure}

\end{document}

\begin{center}

{\Large Supporting Online Material: Strong coupling between single
  electron tunneling and nano-mechanical motion}

G. A. Steele$^1$, 
A. K. H\"{u}ttel$^{1,*}$\symbolfootnote[0]{
$^*$Present address: Institute for Experimental
  and Applied Physics, University of Regensburg, 93040 Regensburg, Germany},
B. Witkamp$^1$, M. Poot$^1$, H. B. Meerwaldt$^1$,
L. P. Kouwenhoven$^1$ \\ and H. S. J. van der Zant$^1$ 

{\em $^1$Kavli Institute of NanoScience, Delft University of Technology, PO Box 
5046, 2600 GA, Delft, The Netherlands.}

\end{center}

\section{Materials and Methods}

\subsection{Device description}

The fabrication is described detail in \cite{Steele2009Tunable} and
\cite{Huettel2009Carbon}: briefly, a trench in a silicon oxide layer
is defined by dry etching, and W/Pt electrodes are deposited to act as
source and drain contacts for injecting current. In device 1, the 
oxide is then further wet-etched by approximately 100 nm to ensure the
nanotube is completely suspended. The nanotube is grown in the last
step of the fabrication to ensure the nanotube is not damaged by
electron beam irradiation or contaminated with residue from chemical
processing.

The length of the trench in device 1 is 800 nm.  The nanotube in
device 1 has a large bandgap of $\sim$300 mV, estimated from the gate
voltage range of depletion of carriers seen in transport measurements,
and a diameter of $\sim$3 nm, determined from the orbital magnetic
moment of the nanotube.  The quantum dot displays a clean four fold
shell filling pattern over a wide range of gate voltages corresponding
to hundreds of Coulomb peaks. In device 1, we drive mechanical
oscillations of the devices using an electric field from a nearby
antenna consisting of an unterminated coaxial cable placed $\sim$2 cm
from the sample. The mechanical nature of the resonance is confirmed
by tuning the nanotube tension using the
gate \cite{Huettel2009Carbon,Sazonova2004Tunable,Witkamp2006BendingMode}.

In device 2, the dry etch is aligned to the source and contacts, and
so a second wet etch is not required. The nanotube in device 2 has a
bandgap of $\sim$100 mV, determined from the size of the empty dot Coulomb
diamond \cite{Steele2009Tunable}. The trench in device 2 has a length
of 430 nm. For device 2, we did not have a high frequency coax for
exciting the mechanical resonance: however, we estimate the resonance
frequency to be in the range of 100 to 500 MHz, based on the length of
the suspended segment.

We have cooled down $\sim$12 suspended nanotube devices made using
this ultra-clean fabrication technique, with trenches ranging from 430
nm to 1100 nm. All of these devices displayed the switch-ridges in
$dI/dV$ discussed in Fig.\ 4 of the main text.

\subsection{Normalization of the resonance signal}

\begin{figure}
\begin{center}
\includegraphics[width=6.5in]{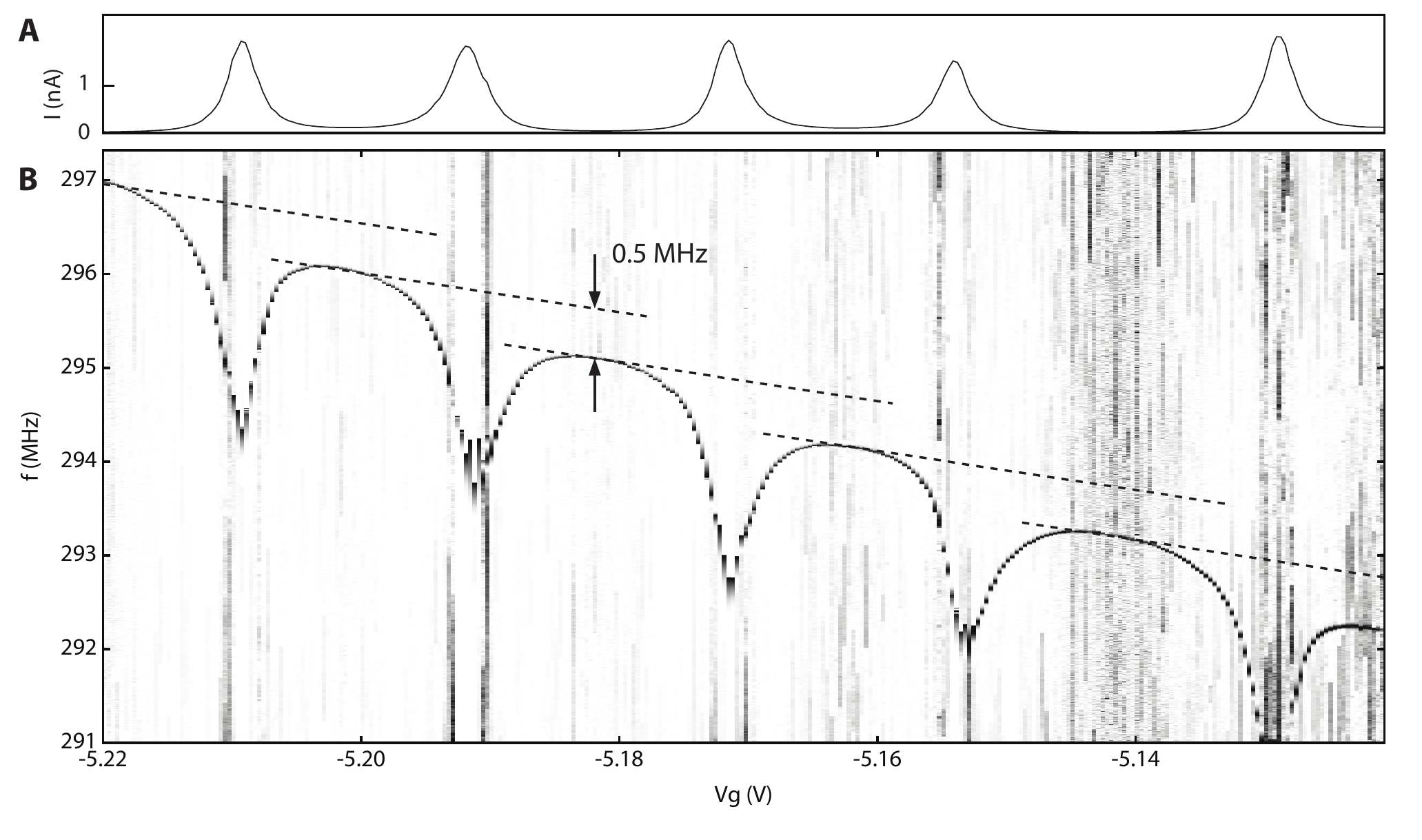}
\end{center}
\caption{(A) Coulomb peaks of the nanotube at $V_{sd} = 0.1$ mV and
  $V_g \sim -5$V. (B) Normalized resonance signal $\Delta I / \Delta
  I_{peak}$, measured at an RF power of -38 dB. We observe shifts of the
  resonator frequency by 0.5 MHz due to the electrostatic force from a
  single electron.}\label{fig:steps}
\end{figure}

As the shift $\Delta I$ of the d.c.\ current from the mechanical
resonance is proportional to the curvature of the Coulomb
peak \cite{Huettel2009Carbon}, it changes in sign and significantly in
magnitude at different gate voltages. In order to clearly show the
position of the resonance in Fig.\ 2 and Fig.\ \ref{fig:steps} at all
gate voltages, we normalize the frequency sweep at each gate voltage
by subtracting the average off-resonant current, taking the absolute
value, and then normalizing to a range of 0 to 1.

\section{Supplementary Text}

\subsection{Single electron tuning shifts of 0.5 MHz}

In Fig.\ 2 of the main text, we showed offsets of the resonator
frequency of 0.1 MHz due to the static charge of a single electron.
The data from Fig.\ 2 of the text was taken around $V_g \sim -1$V.  In
Fig.\ \ref{fig:steps}, we show the resonance frequency as a function of
gate voltage around $V_g \sim -5$V. Here, we observe offsets of the
resonator frequency of 0.5 MHz.  At larger gate voltages, the
frequency tuning curve shown in Fig.\ 1C of the main text becomes
steeper, and thus the offset in frequency from the single electron
charge becomes larger.  Although the nanotube quantum dot charge
still increases by only one electron, this electron exerts a larger
force on the nanotube due to its stronger attraction to the larger
total charge on the gate.

\subsection{Frequency softening by Coulomb blockade} \label{sec:soft}

\begin{figure}
\begin{center}
\includegraphics[width=5.5in]{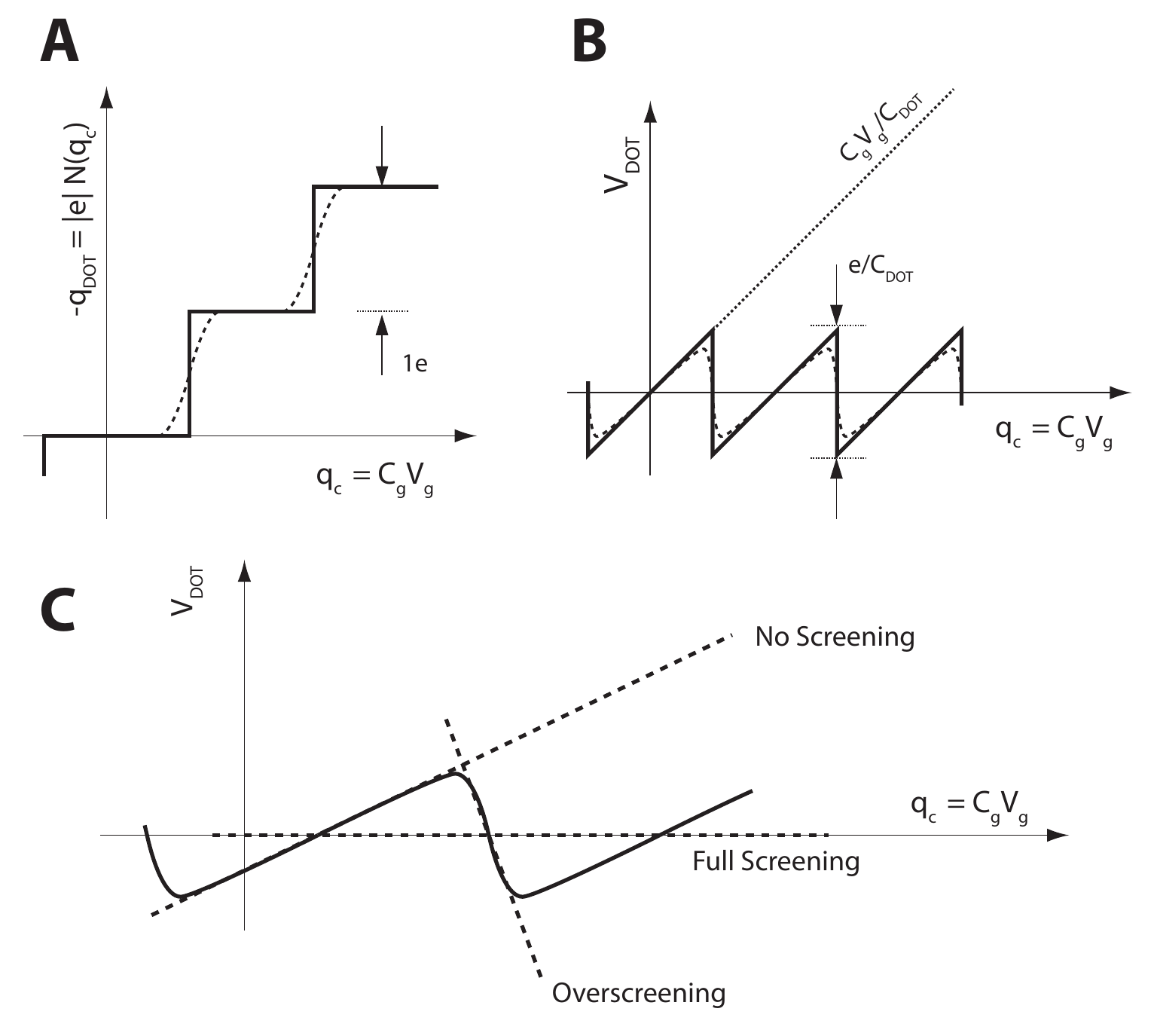}
\end{center}
\caption{(A) The charge on the quantum dot $q_{dot}$ as a function of
  the control charge $q_c = C_g V_g$ follows a staircase with steps of
  $e$. The dashed line shows the average dot occupation $\langle
  N(q_c) \rangle$: the steps become broadened by charge
  fluctuations. (B) The potential of the quantum dot $V_{dot}$
  oscillates with a sawtooth waveform. On the rising edge, $V_{dot}$
  follows the potential of a floating conductor, $C_g V_g / C_{dot}$
  (dotted line). Averaging over charge fluctuations, the sharp edges
  of the sawtooth become broadened ($\langle V_{dot} \rangle$ shown by
  dashed line). (C) shows the time averaged $\langle V_{dot}
  \rangle$ as a function of the control charge. Between charge
  transitions, the quantum dot charge is fixed, and does not screen
  $V_g$ at all (no screening). At charge transitions, $V_{dot}$
  drops suddenly, overscreening $V_g$. On average, $V_{dot}$ is
  constant (full screening).}\label{fig:cb}
\end{figure}

The dips in the frequency of the mechanical oscillator, shown in
Figs. 2 and 3 of the main text, arise from a softening of the
electrostatic component of the spring constant of the mechanical
motion due to Coulomb blockade \cite{brinkthesis}. To calculate this
softening, we begin with the electrostatic force between the dot and
the gate, given by
\begin{equation} \label{eq:force}
F_{dot} = \frac{1}{2} \frac{dC_g}{dz}(V_g - V_{dot})^2
\end{equation}
where $C_g$ is the capacitance of the quantum dot to the gate,
$V_{dot}$ is the electrostatic potential of the quantum dot, and $z$
is the vertical distance between the nanotube and the quantum
dot. Since $V_g$ is fixed and $dC_g/dz$ is slowly varying, changes in
this force will be dominated by changes in $V_{dot}$. The voltage on
the dot is found from electrostatics:
\begin{equation} \label{eq:vdot}
V_{dot} = \frac{C_g V_g + q_{dot}}{C_{dot}}
\end{equation}
where $C_{dot}$ is the total capacitance of the quantum dot, and
$q_{dot}$ is the charge on the quantum dot. In Coulomb blockade, the
charge on the dot does not increase continuously with gate voltage,
but instead increase in discrete steps of one electron, as illustrated
in Fig.\ \ref{fig:cb}A. As a result, the electrostatic potential on the dot
will oscillate in a sawtooth pattern with an amplitude of $e/C_{dot}$,
as shown in Fig.\ \ref{fig:cb}A. We write the charge on the dot as:
\begin{equation} \label{eq:qdot}
q_{dot} =  -|e| N(q_c)
\end{equation}
where $N$ is the average number of electrons on the quantum dot, and
$q_c = C_g V_g$ is the ``control charge''. Note that the control
charge is not the charge on the gate: the charge on the gate is given
by the voltage difference from the gate to the dot, $q_g = C_g (V_g -
V_{dot})$. The control charge is a concept we use here to express the
idea that the quantum dot is controlled by both the voltage on the
gate $V_g$ and by the (distance dependent) capacitance to the gate
$C_g(z)$. The control charge is the continuous charge that would be on
the quantum dot in the absence of Coulomb blockade ($V_{dot} = 0$). 

In the ideal limit of zero temperature and opaque tunnel barriers, $N$
would follow a staircase with sharp steps (Fig.\ \ref{fig:cb}A solid
line), and the sawtooth oscillation of $V_{dot}$ would have sharp
edges (Fig.\ \ref{fig:cb}B solid line). In practice, however, the
electron number near the transition will fluctuate in time due to
finite temperature and tunnel coupling to the leads. The transitions
in the average charge $\langle N \rangle$ are then smooth, acquiring a
finite width (dashed lines in Fig.\ \ref{fig:cb}A and B). The timescale
of these fluctuations is set by the tunneling time $\Gamma^{-1}$. If
the mechanical motion is much slower than $\Gamma$, the resonator will
feel a force averaged over these fluctuations, which can be calculated
using the time averaged expression for $\langle N(q_c) \rangle$. To
find correction to the mechanical spring constant $\Delta k$ from this
force, we take the derivative of Eq.\ \ref{eq:force} with respect to the
displacement of the nanotube:
\begin{equation}
\Delta k = - \frac{dF_{dot}}{dz} =  (V_g - V_{dot}) \frac{dC}{dz}
\frac{dV_{dot}}{dz}
\end{equation}
(neglecting slowly varying terms $d^2C_g/dz^2$). Using
Eq.\ \ref{eq:vdot} and \ref{eq:qdot}, we obtain:
\begin{equation}\label{eq:k}
\Delta k = \frac{V_g (V_g - V_{dot})} {C_{dot}}
\left( \frac{dC_g}{dz} \right)^2 \left(1 - |e| \frac{d\langle N \rangle}{dq_c}\right)
\end{equation}
Note that the second term in the brackets on the right leads to a
softening of the spring constant of the resonator, proportional to how
quickly the average quantum dot occupation changes through a charge
transition (i.e. the Coulomb peak). The strong peak in $d\langle
N\rangle /dq_c$ at the steps of the Coulomb staircase leads to the
large dips in the resonance frequency we observe in Figs. 2 and 3 in
the main text. The softening that arises from the negative sign in
front of $d\langle N\rangle /dq_c$ in Eq.\ \ref{eq:k} is very
non-intuitive: we expect that increasing the charge on the nanotube
will, in general, pull the nanotube towards the gate, increasing the
tension and stiffening the spring constant. Here, increasing the
charge instead {\em softens} the spring constant.

Physically, this softening comes from the peculiar screening
properties of a Coulomb blockaded quantum dot, illustrated in Fig.\
\ref{fig:cb}c. Between charge transitions, the quantum dot acts like a
floating island that does not screen the gate potential at all
($dV_{dot}/dV_g > 0$). At the charge transition, the quantum dot
compensates by {\em overscreening} the gate potential giving a sudden
drop in the dot potential ($dV_{dot}/dV_g < 0$). Over many charge
transitions, the net effect is that the average dot potential stays
fixed \cite{quantum}, as is the case for full screening by a metal
conductor ($dV_{dot}/dV_g = 0$). It is the overscreening at the
negative steps in the dot potential that leads to the softening of the
spring constant.

In Fig.\ 2C of the main text, we demonstrate that the frequency dip
becomes broader and shallower with increasing bias across the
dot. This can be understood from the effect of finite bias on the
average dot occupation $\langle N(q_c) \rangle$: at large bias, the
average occupation of the quantum dot changes more slowly as its
chemical potential moves through the bias window. The derivative $d\langle
N\rangle /dq_c$ is smaller, and hence the shift of the spring constant
(Eq.\ \ref{eq:k}) is also smaller. It is also interesting to note that
for asymmetric tunnel barriers ($\Gamma_L \neq \Gamma_R$), the quantum
dot occupation at higher bias changes more quickly at one edge of the
Coulomb diamond, whereas the current does not. The peak in $d\langle
N\rangle /dq_c$ therefore does not have to coincide with the maximum
current of the Coulomb peak. This is also observed in the
measurements, and can be seen clearly, for example, in
Fig.\ \ref{fig:nl}.

\subsection{Coulomb blockade induced nonlinearity of the resonator}

\begin{figure}
\begin{center}
\includegraphics[width=6in]{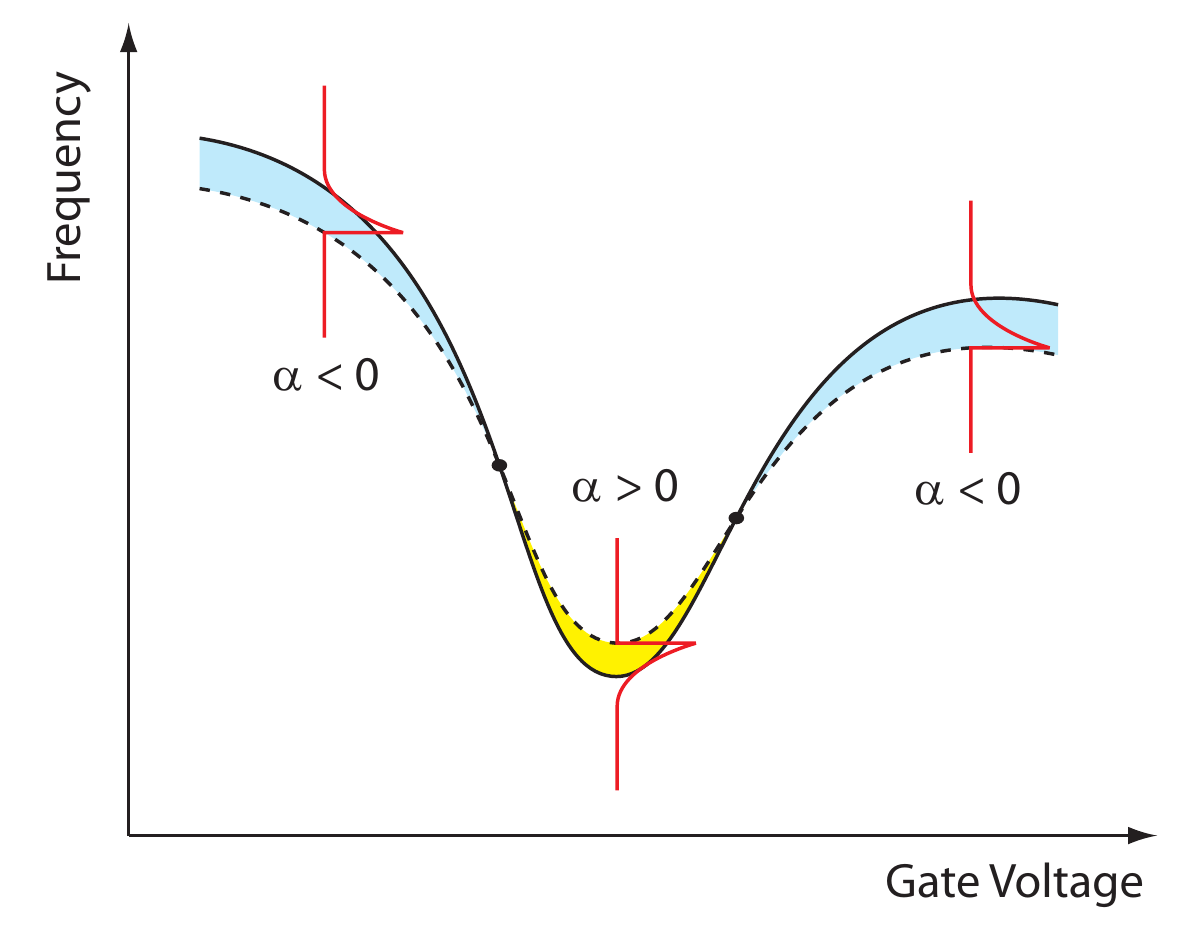}
\end{center}
\caption{ The sign of the nonlinear Duffing parameter $\alpha$ is
  determined by the curvature of $f_0(V_g)$.  The solid line shows the
  position of the resonance at low powers, $f_0$. The dashed line
  shows the sharp edge of the non-linear resonance lineshape at higher
  powers (frequency traces illustrated by red lines). For positive
  curvature (yellow), we have $\alpha > 0$, and for negative curvature
  (blue), we have $\alpha < 0$.}\label{fig:alpha}
\end{figure}

\begin{figure}
\begin{center}
\includegraphics[width=6in]{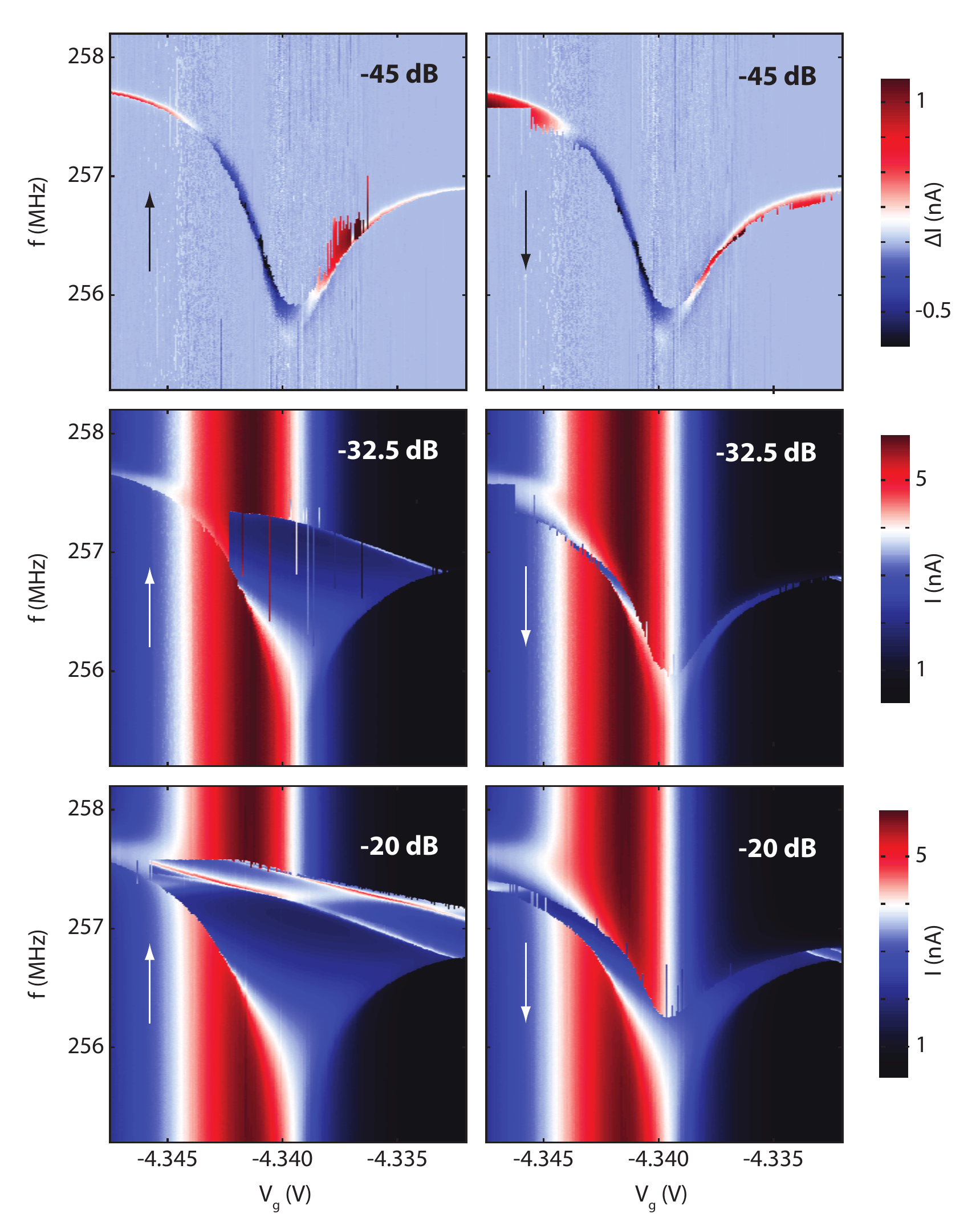}
\end{center}
\caption{ Upwards (left column) and downwards (right column) frequency
  sweeps in the non-linear regime, taken at -45 dB, -32.5 dB and -20
  dB for the same $V_{sd} = 0.5$ mV and $V_g$ range as used in figure
  3 of the main text. For the -45 dB data, the off-resonant current
  was subtracted from the data.}\label{fig:nl}
\end{figure}

In section \ref{sec:soft}, we calculated the change in the linear
spring constant of the mechanical resonator due to the force on the
resonator from the Coulomb blockaded quantum dot. In addition to a
correction to the linear coefficient, there will also be terms higher
order in the displacement $dz$ from Eq.\ \ref{eq:force}. The Duffing
parameter $\alpha$, which determines the initial softening or
hardening spring behaviour, can be calculated by taking the third
derivative of the force $d^3F/dz^3$. However, as we can read off
$\Delta k(V_g)$ directly from the measured gate voltage dependence of
the low-power resonance frequency, $f_0(V_g)$, we can also predict
$\alpha$ from the experimental data:
\begin{equation}
\alpha = - \frac{d^3F}{dz^3} = \frac{d^2}{dz^2}\ \Delta k (q_c) =
V_g^2 \left( \frac{dC}{dz} \right)^2 \frac{d^2(\Delta k)}{dq_c^2},
\end{equation}
again neglecting terms proportional to $d^2C/dz^2$. The sign of
$\alpha$ will follow the sign of the curvature of $\Delta k(V_g)$, as
determined from the observed $f_0(V_g)$. This gives a change in the
sign of $\alpha$ at the inflection points of the frequency dip, as
illustrated in figure \ref{fig:alpha}.

From the mechanical deformation of a beam under tension, we would
normally expect a hardening spring behavior, as observed in previous
nanotube experiments \cite{Sazonova2004Tunable}. (This can also be seen
from the overall positive curvature of the mechanical tuning of the
resonance, shown in Fig.\ 1D of the main text.) The fact that we
observe both softening and hardening behaviour with a small change in
gate voltage indicates that the nonlinear coefficient from the single
electron force, $\alpha_{e}$, is much larger in magnitude than that
from the mechanical deformation, $\alpha_{mech}$: $|\alpha_{e}| \gg
|\alpha_{mech}|$. Essentially, the single electron force dominates the
nonlinearity of the resonator.

\subsection{Non-linear behaviour at high driving powers}

\begin{figure}
\begin{center}
\includegraphics[width=6.5in]{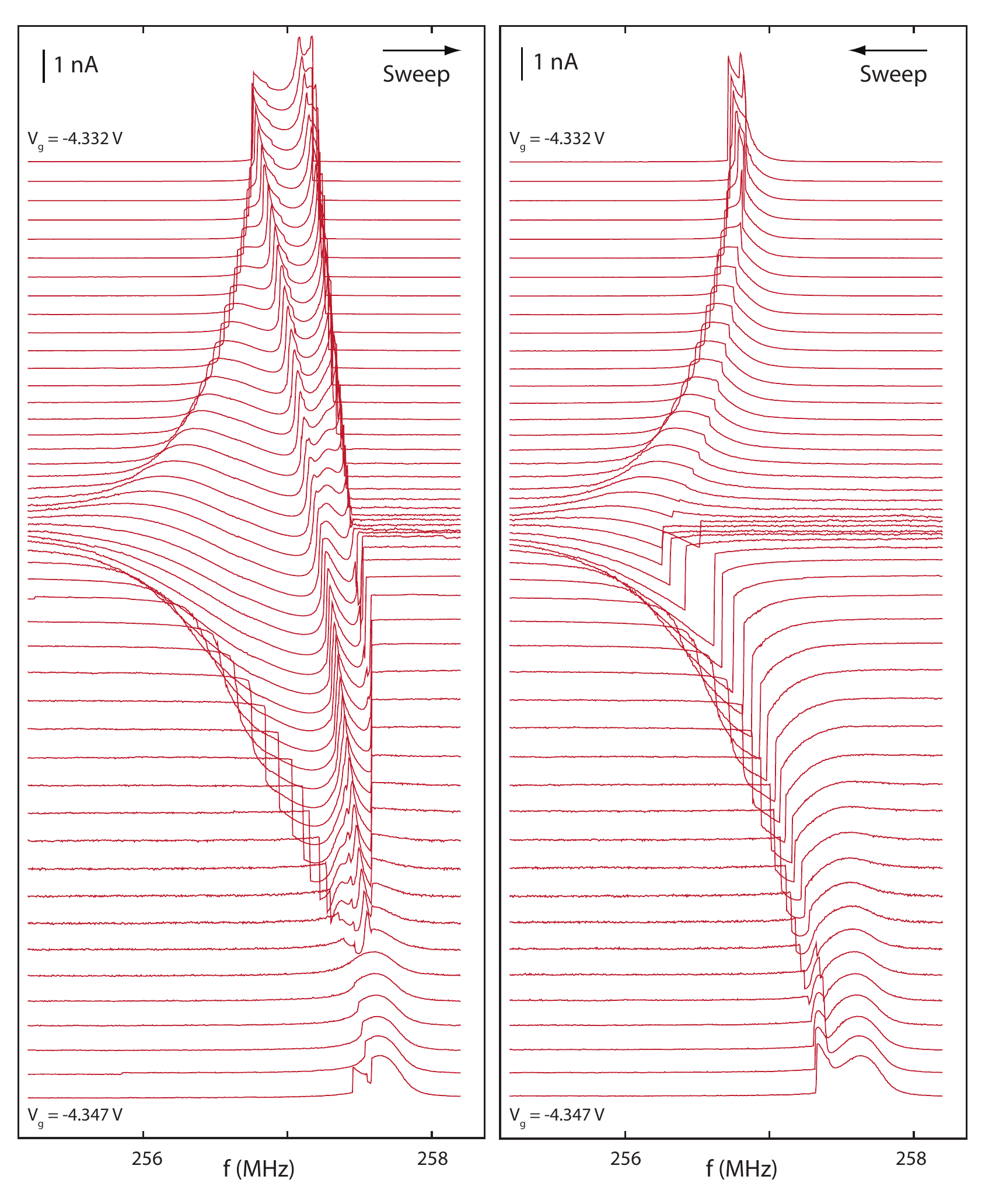}
\end{center}
\caption{ A waterfall plot of upward (left) and downward (right)
  frequency sweeps at a power of -20 dB. Line cuts are data from the
  colorscale plots displayed in Fig.\ \ref{fig:nl}.}\label{fig:water}
\end{figure}

Fig.\ \ref{fig:nl} shows the data from Fig.\ 3 of the main text in
both upwards and downwards frequency sweep directions, and as well for
a power of -32.5 dB. The resonance at the highest powers displays a
highly structured lineshape, particularly around the frequency dip, as
can be seen in the waterfall plot (Fig.\ \ref{fig:water}) of the -20
dB data.

\subsection{Suppression of switch ridges at low tunnel rates}

In figure 4 of the main text, we present a peculiar instability in the
Coulomb diamonds of clean, suspended carbon nanotube quantum
dots. This instability appeared as ridges of sharp positive and
negative spikes in the differential conductance, visible also as a
sudden jump in the current. Unlike electronic excited
states \cite{Foxman1993Effects} or phonon
sidebands \cite{Park2000Nanomechanical,
Sapmaz2006Tunneling,Zwanenburg2009Spin,Leturcq2009FranckCondon}, the
ridges of spikes do not run parallel to the Coulomb diamond edges, nor
are they broadened by temperature or tunneling rates as the electronic
excited states are: they often occur over just one pixel in the
measurement.  The spikes arise from a positive feedback mechanism
between single electron tunneling and the mechanical motion that
spontaneously drives the resonator into a high amplitude oscillation
state \cite{Usmani2007Strong}. The feedback mechanism in
\cite{Usmani2007Strong} requires a mechanical resonator with a very
high quality-factor, a quantum dot with energy dependent tunneling,
and a tunneling rate of electrons through the dot that is much faster
than the mechanical resonance frequency. Essentially, for each
mechanical oscillation of the nanotube, many electrons should tunnel
through the quantum dot.

\begin{figure}
\begin{center}
\includegraphics[width=6.5in]{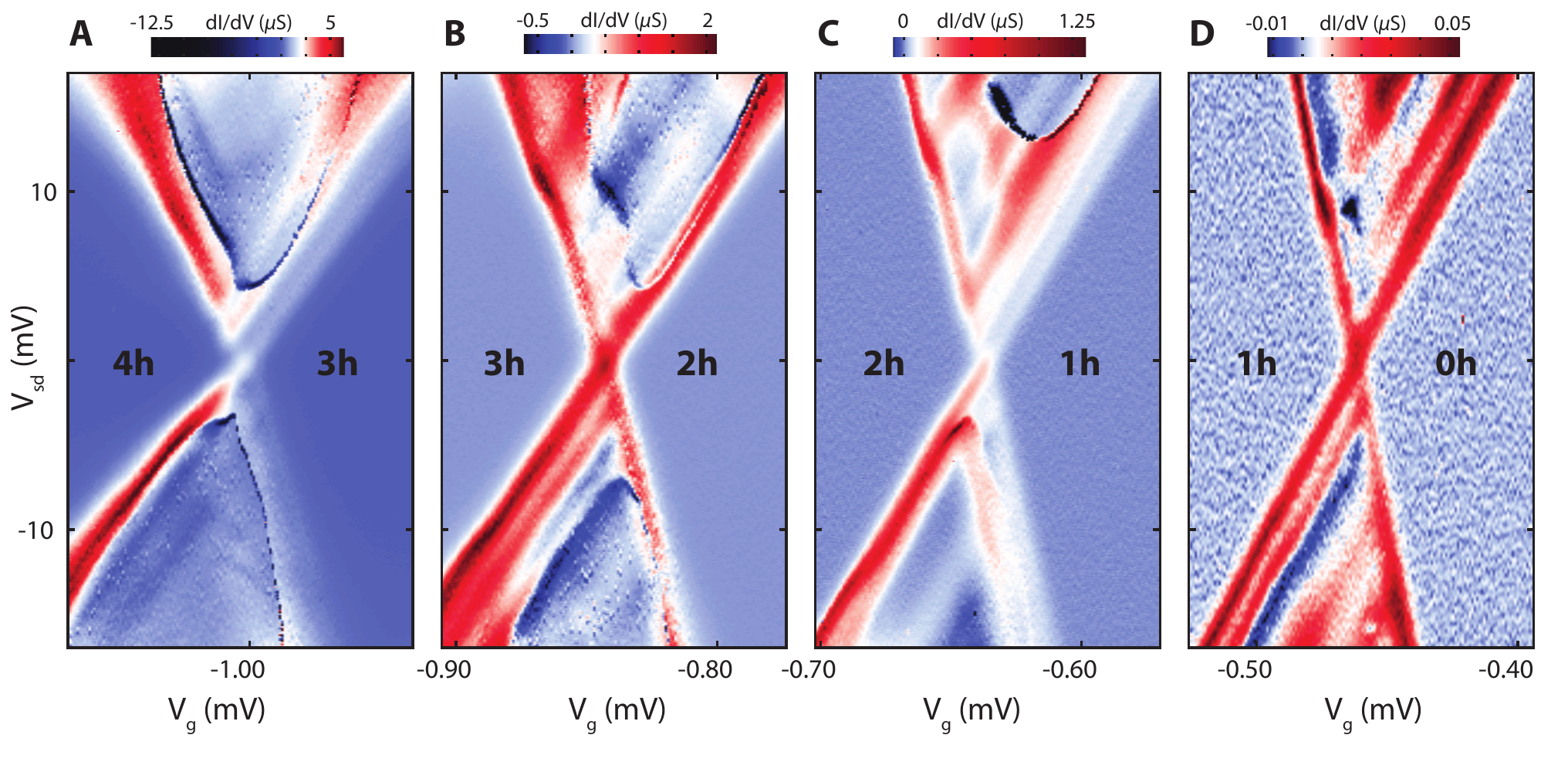}
\end{center}
\caption{ Suppression of the mechanical instability by reducing tunnel
  rates using tunable barriers in device 2. (A) The data from figure
  4(B) of the main text, showing spikes in the differential
  conductance $dI/dV$. (B) - (D) Reducing the number of holes in the
  quantum dot to zero, the tunnel barriers become more opaque. The
  spike ridges in the differential are initially pushed to higher
  bias, and then suppressed in the last Coulomb diamond. For a
  comparison, the current at a 4 mV bias in (A) - (D) is 7.3 nA, 3.1
  nA, 0.5 nA, and 80 pA, corresponding to tunnel rates $\Gamma$ of 45
  GHz, 20 GHz, 3 GHz, and 500 MHz, respectively.}\label{fig:sw}
\end{figure}

In Fig.\ \ref{fig:sw}, we show that, experimentally, this instability
can be suppressed by reducing the rate at which electrons tunnel
through the dot. This data is from device 2, in which the barriers to
the leads changed very rapidly as we reduced the number of holes in
the quantum dot. Although we did not have RF coax for measuring the
mechanical resonance frequency of this device, we estimate the
resonance frequency to be on the order of 100 to 500 MHz from the
length of the nanotube. At high tunneling rates, corresponding to
$\Gamma \gg f_0$, Coulomb diamonds all display ridges of spikes in
$dI/dV$. In Fig.\ \ref{fig:sw}D, the dot is very weakly coupled to the
leads, with a tunnel rate of 500 MHz at $V_{sd}$ = 4 mV. $\Gamma$ is
no longer much larger than $f_0$, and the instability is suppressed.

%\pagebreak 

\section{Supplementary Figures}

\ \\

\section{Supplementary References}

\vspace{-0.6in}

\renewcommand{\refname}{}

%\bibliography{paper}{}